\documentclass{cseg2015}
\usepackage{xy}
\usepackage{amsmath,amssymb,graphicx,natbib}
\usepackage{rotating}
\usepackage{lscape}
\definecolor{burntorange}{RGB}{225,100,0}
\usepackage{setspace}
\usepackage{wrapfig}

\begin{document}
 \vspace*{-2.cm}
\begin{center}
\includegraphics[height=3.2cm]{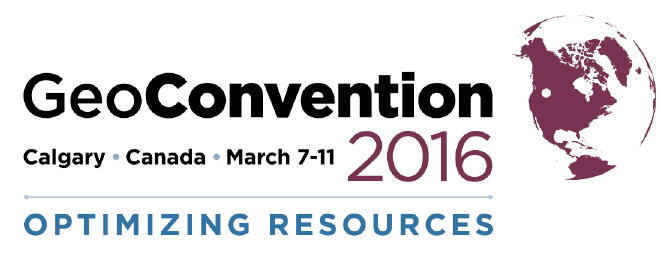}
\end{center}

{\LARGE {\bf Application of shifted-Laplace preconditioners for heterogenous Helmholtz equation{--} part 1: Data modelling}} 


 \vspace*{-0.1cm}

{\it Nasser Kazemi}

\vspace*{-0.2cm}

{\it Department of Physics, University of Alberta}

\section{Summary}

In several geophysical applications, such as full waveform inversion and data modelling, we are facing the solution of inhomogeneous Helmholtz equation. The difficulties of solving the Helmholtz equation are two fold. Firstly, in the case of large scale problems we cannot calculate the inverse of the Helmholtz operator directly. Hence, iterative algorithms should be implemented. Secondly, the Helmholtz operator is non-unitary and non-diagonalizable which in turn  deteriorates the performances of the iterative algorithms (especially for high wavenumbers). To overcome this issue, we need to implement proper preconditioners for a Krylov subspace method to solve the problem efficiently. In this paper we incorporated shifted-Laplace operators to precondition the system of equations and then generalized minimal residual (GMRES) method used to solve the problem iteratively. The numerical results show the performance of the preconditioning operator in improving the convergence rate of the GMRES algorithm for data modelling case. In the companion paper we discussed the application of preconditioned data modelling algorithm in the context of frequency domain full waveform inversion. However, the analysis of the degree of suitability of the preconditioners in the solution of Helmholtz equation is an ongoing field of study.
\section{Introduction}
Solving time-harmonic wave equations can result in a linear system with a non-normal data matrix. The non-normality of the data matrix is due to the losses and boundary conditions. For large scale problems, application of iterative methods is necessary. Convergence of Krylov subspace algorithms for the linear system of equations with normal data matrices can be explained with the eigenvalue spectrum of the data matrix. However, in the case of non-normal data matrix, the convergence properties of the system is a bit more complicated. \cite{Gijzen} use the eigenvalue spectrum of the preconditioned Helmholtz operator to derive the upper bounds on the convergence of GMRES \cite[]{GMRES} method and they choose optimal shifting parameter for the preconditioner that minimizes the estimated upper bound. Unfortunately, these calculations are not valid if the circle around the eigenvalues encloses origin. 

On the other hand, one can use pseudo spectrum of Helmholtz operator to calculate the worst case convergence behaviour of Krylov subspace algorithms \cite[]{Trefethen,Sifuentes,Hannukainen}. Pseudo spectrum is non-convex and can curl around the origin. Preconditioners will change the shape and extent of the bounding regions of the pseudo spectrum of the conditioned data matrix, hence the convergence rate. The convergence rate of the GMRES method for the solution of the preconditioned system is independent of the mesh size which in turn prevents extensive growth of the size of data matrix for high wavenumbers. However, the number of iterations required to reach a predefined accuracy is dependent on the wavenumber \cite[]{Ernst,Hannukainen}. In this paper we will show numerically, how shifted-Laplace preconditioners can change the convergence rate of the GMRES algorithm for the solution of heterogenous Helmholtz problem. \cite {kazemi-fwi} also showed the convergence improvement of frequency domain full waveform inversion technique after using the preconditioned data modelling algorithm presented in here.
 
\section{Data modelling}
The Helmholtz equation with first-order absorbing boundary condition can be written as
\begin{equation}\label{eq:1}
 -\nabla . \rho \nabla {\bf u}-{k}^2(1-\hat{j} \alpha) {\bf u}={\bf f} \quad \mbox{in}\quad \Omega \in {\mathbb R}^d\quad d=1,2,3
 \end{equation}
where $\rho$ is density, ${k({\bf x},{\bf z})=\omega/c({\bf x},{\bf z})}$ is wavenumber and varies within $\Omega$ due to spatial variation of velocity field $c({\bf x},{\bf z})$, $\omega$ is angular frequency related to the source function ${\bf f}$, $\alpha$ is damping parameter, ${\bf u}$ is pressure wave-field and ${\hat{j}=\sqrt -1}$ is complex identity. For constant density, equation (\ref{eq:1}) changes to
\begin{equation}\label{eq:2}
 -\Delta {\bf u}-\omega^{2}{\bf m}^{2}(1-\hat{j} \alpha) {\bf u}={\bf f}
 \end{equation}
where ${\bf m}=c({\bf x},{\bf z})^{-1}$ is slowness. In matrix-vector notation equation (\ref{eq:2}) can be expressed as
\begin{equation}\label{eq:3}
{\bf H} {\bf u}={\bf f}
 \end{equation}
where {\bf H} is data matrix and usually indefinite and non-normal. There is a possibility of solving equation (\ref{eq:3}) via direct methods such as LU factorization for small scale problems;
 \begin{equation}\label{eq:4}
\hat {\bf u}={\bf H}^{-1} {\bf f}.
 \end{equation}
\begin{figure}[h]
  \vspace{-0.1cm}
  \begin{center}
    \includegraphics[width=0.5\textwidth,height=10cm]{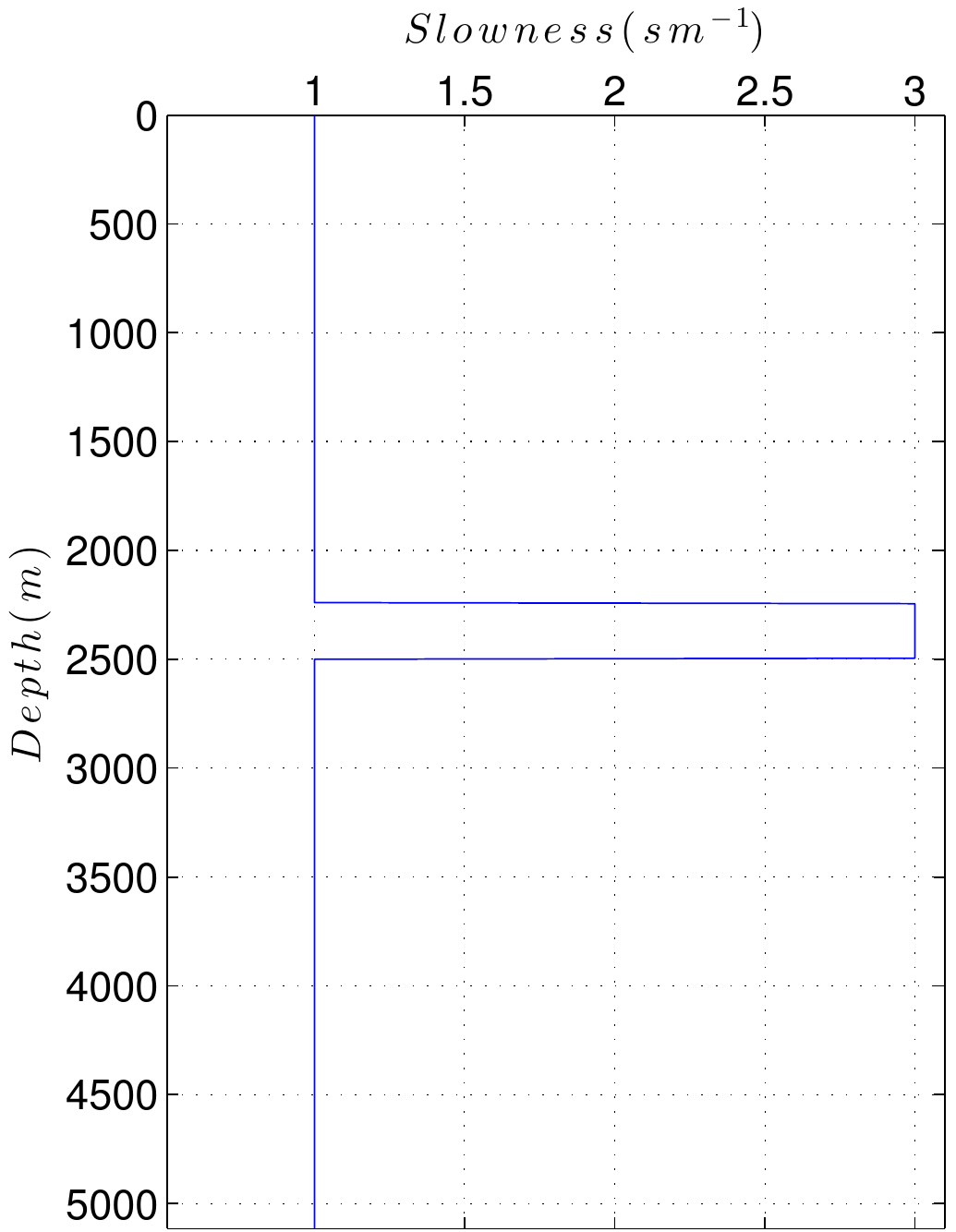}
  \end{center}
  \caption{True 1D scattering wavefield model. }
  \label{fig:1}
  \end{figure}

Nevertheless, in geophysical applications, computation of the inverse of the ${\bf H}$ matrix is a difficult task \cite[]{Ernst}. This is because the data matrix is too big, indefinite and non-normal. Hence, utilization of the iterative algorithms seems necessary. In this paper we will use GMRES method. However, the performances of the Krylov subspace approach are dependent on the behaviour of the data matrices. ${\bf H}$ matrix can be changed to some well-behaved ones via proper preconditioners. This in turn will raise the question on which kind of preconditoners should be chosen and implemented. Analysis of the degree of suitability of the preconditioners for Helmholtz operator is complex. Here, we only discuss shifted-Laplace preconditioners. Given a matrix ${\bf M}={\bf M}_l\; {\bf M}_r\; \in \mathbb {C}^{N\times N}$, preconditioned Helmholtz equation for constant density case reads
\begin{equation}\label{eq:5}
{\bf M}_l^{-1}\;{\bf H}\; {\bf M}_r^{-1}\;{\bf z}={\bf M}_l^{-1}\; {\bf f} \quad \mbox{and}\quad {\bf z}={\bf M}_r\;{\bf u}
 \end{equation}
where ${\bf M}_l$ and ${\bf M}_r$ are left and right preconditioners, respectivily. We set
\begin{equation}\label{eq:6}
{\bf M}_l={\bf I} \quad  \mbox{and} \quad {\bf M}_r={\bf P}= -\Delta + \beta {k}^2
 \end{equation}
 with $\beta=\beta_1+\hat{j}\beta_2 \in \mathbb {C}$ as a complex shifting parameter and ${\bf I}$ is an identity matrix. Hence equation (\ref{eq:5}) simplifies to 
\begin{equation}\label{eq:7}
{\bf H}\; {\bf P}^{-1}\;{\bf z}={\bf f} \quad \mbox{and}\quad {\bf z}={\bf P}\;{\bf u}.
 \end{equation} 

\begin{figure}[]
\vspace{-1cm}
\centerline{\includegraphics[width=0.5\columnwidth,height=10cm]{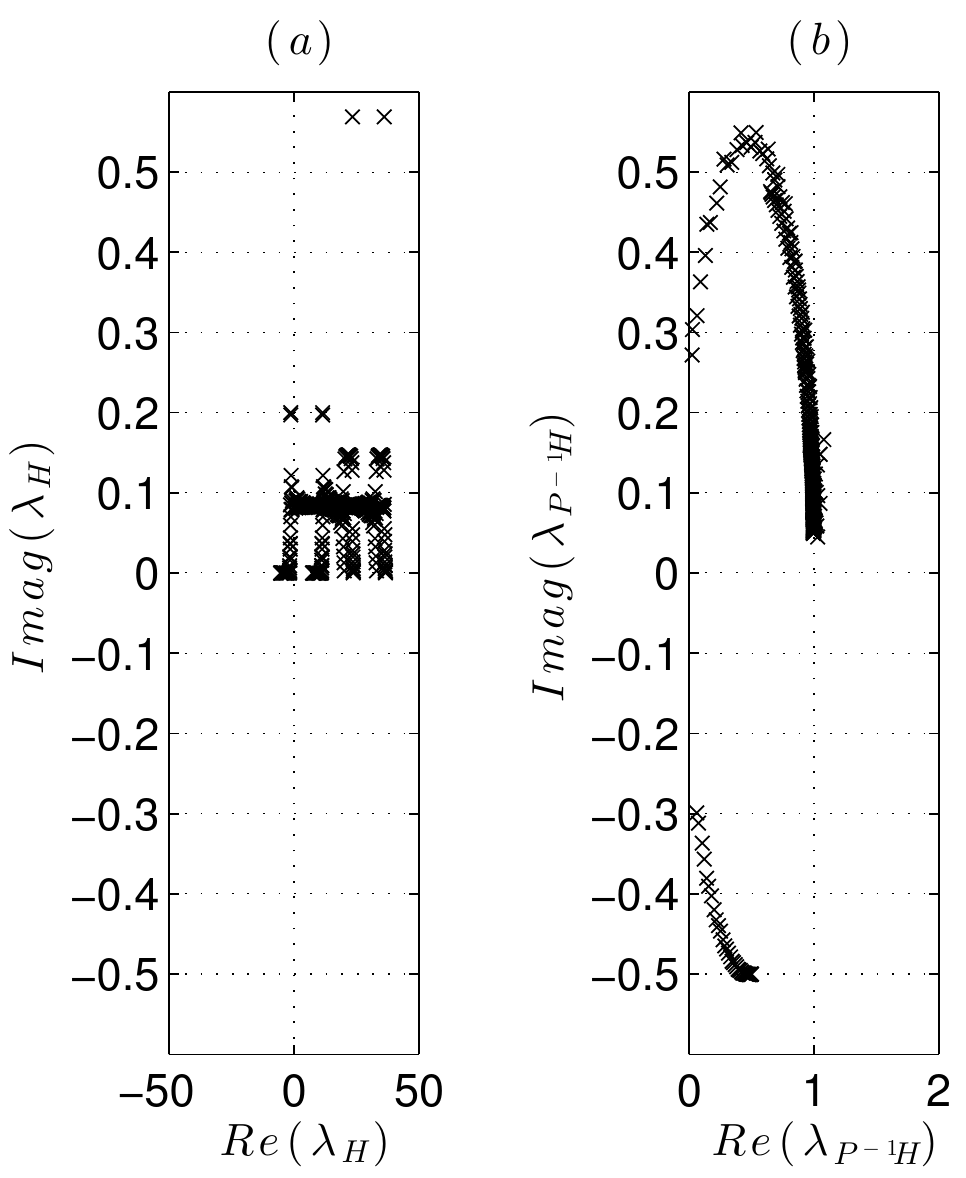}}
\caption{Eigenvalue spectrum of Helmholtz operator corresponding to the model in Figure \ref{fig:1}. }
\label{fig:2}
\end{figure}

 Analysis of the convergence rate of GMRES method for the solution of equations (\ref{eq:7}) and (\ref{eq:3}), due to the indefiniteness and non-normality of the data matrix is not easy and is outside of the scope of this paper. Generally speaking, there is a relationship between the condition number of data matrix and convergence rate of the GMRES algorithm. Hence, we are trying to cluster the eigenvalues of the conditioned system around unit circle. 
 \vspace{+.2cm}
\section{Example}

\begin{figure}[]
\vspace{-1.1cm}
\centerline{\includegraphics[width=1\columnwidth]{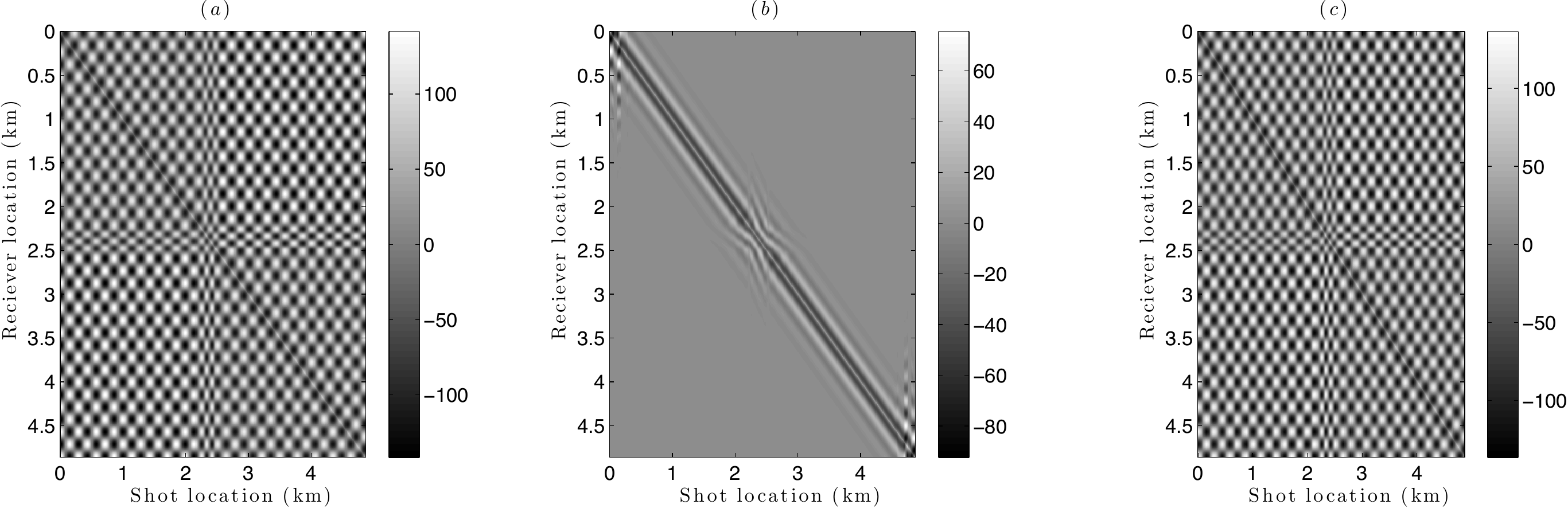}}
\caption{1D modelling results. a) modelling with LU factorization. b) GMRES without preconditioners. c) GMRES with preconditioners.}
\label{fig:3}
\end{figure}

\begin{figure}[]
\vspace{-.4cm}
\centerline{\includegraphics[width=0.5\columnwidth]{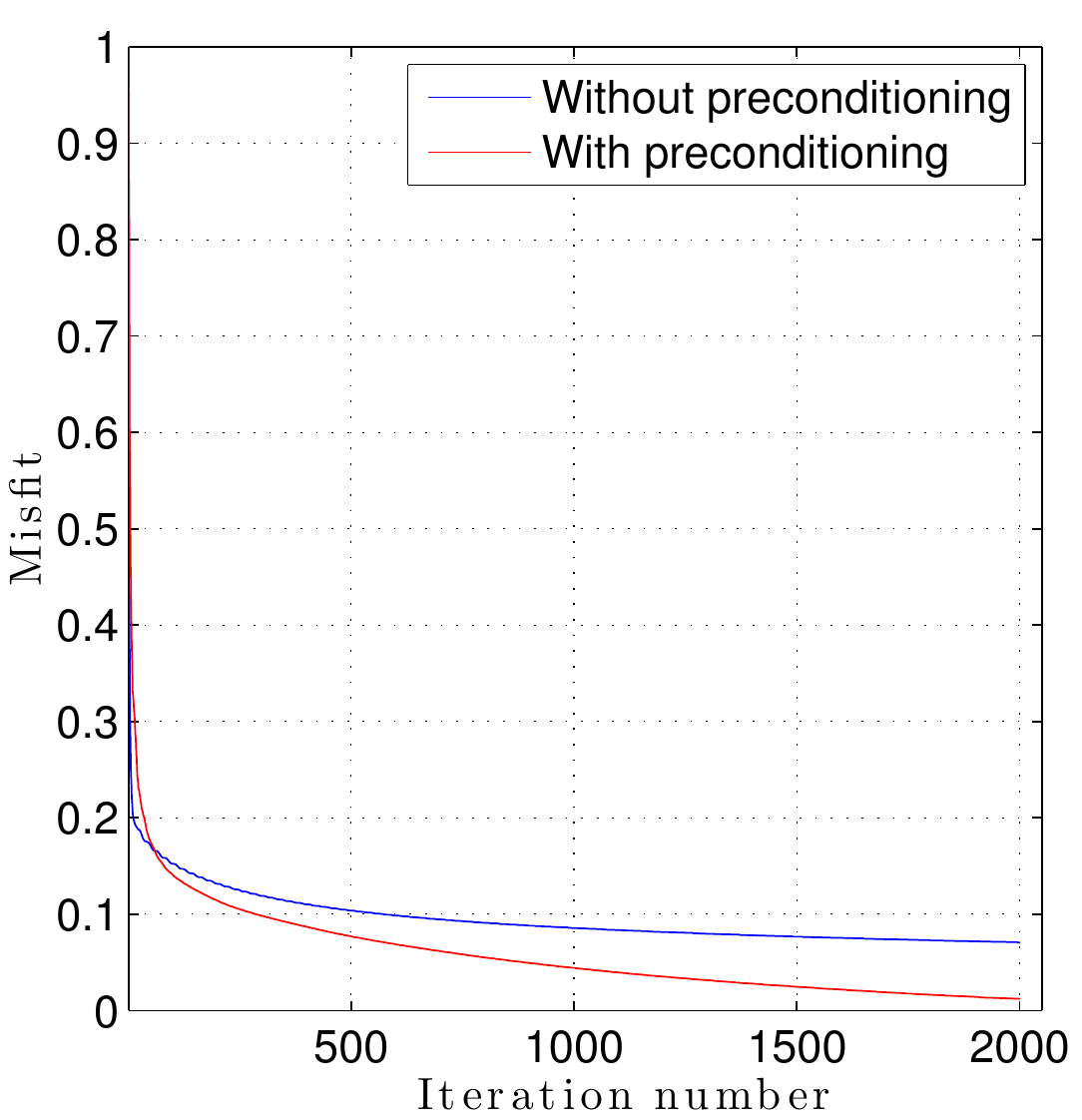}}
\caption{Comparison of convergence rate of GMRES algorithm for the modelled data in Figure \ref{fig:3}.}
\label{fig:4}
\end{figure}

To demonstrate the method we generated a 1D model (Figure \ref{fig:1}). Sources and receivers are placed along the Well. The Helmholtz equation for this geometry are organized in the frequency domain. The eigenvalue spectrum of the Helmholtz operator corresponding to $30 Hz$ is shown in Figure \ref{fig:2}a. The spectrum of the preconditioned operator (i.e., ${\bf P}^{-1}{\bf H}$) is represented in Figure \ref{fig:2}b. As it is clear from the Figure, after application of the shifted-Laplace preconditioner with ($\beta_1=1$ and $\beta_2=1$) we were be able to cluster the eigenvalues of the Helmholtz operator around a unit circle and the eigenvalues are shifted from origin and negative real values toward positive real values. This property in theory should help us to increase the convergence rate of GMRES method. Later on we implemented LU factorization, GMRES without preconditioner and GMRES with preconditioner to model the data (Figure \ref{fig:3}). GMRES algorithm without preconditioners slowly converges and we cannot model the data properly, however, after preconditioning the operator the modelled data is very similar to LU factorization result. The convergence rate of the GMRES method with and without preconditioners is compared in Figure \ref{fig:4}. 

\begin{figure}[]
\vspace{-.7cm}
\centerline{\includegraphics[width=0.75\columnwidth]{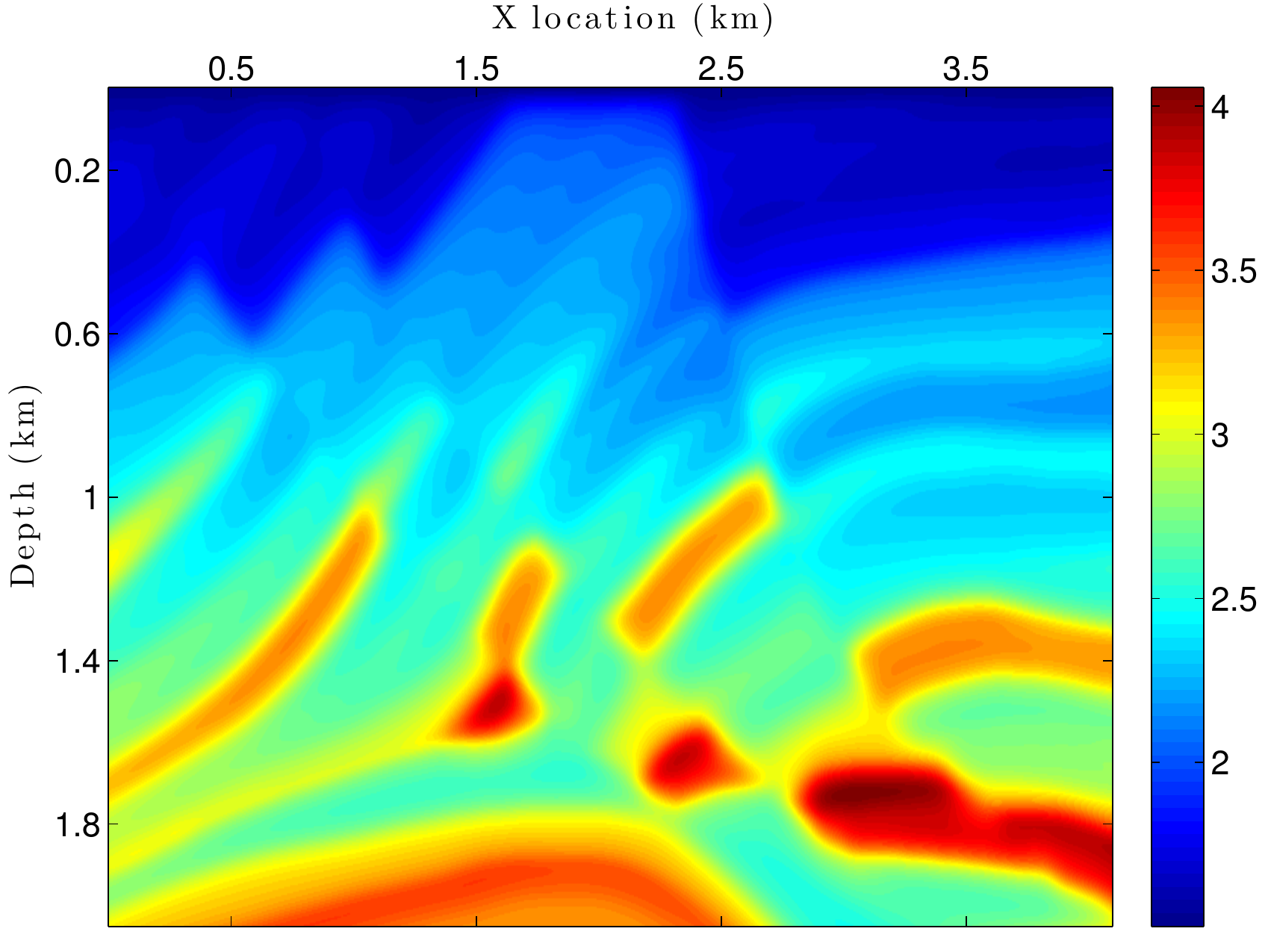}}
\caption{True 2D scattering wavefield model (middle part of Marmousi model).}
\label{fig:5}
\end{figure}

Next we applied this method on Marmousi model. The scattering field (${\bf m}$) is shown in Figure \ref{fig:5}. For this example we only show one monochromatic ($10 Hz$) wavefield corresponding to the shot in the middle. Again, GMRES algorithm without preconditioners cannot model the data properly. These results prove that Krylov subspace vectors do not have appropriate support to approximate the solution and application of the preconditioners is necessary.

\begin{figure}[]
\vspace{-0.5cm}
\centerline{\includegraphics[width=1\columnwidth]{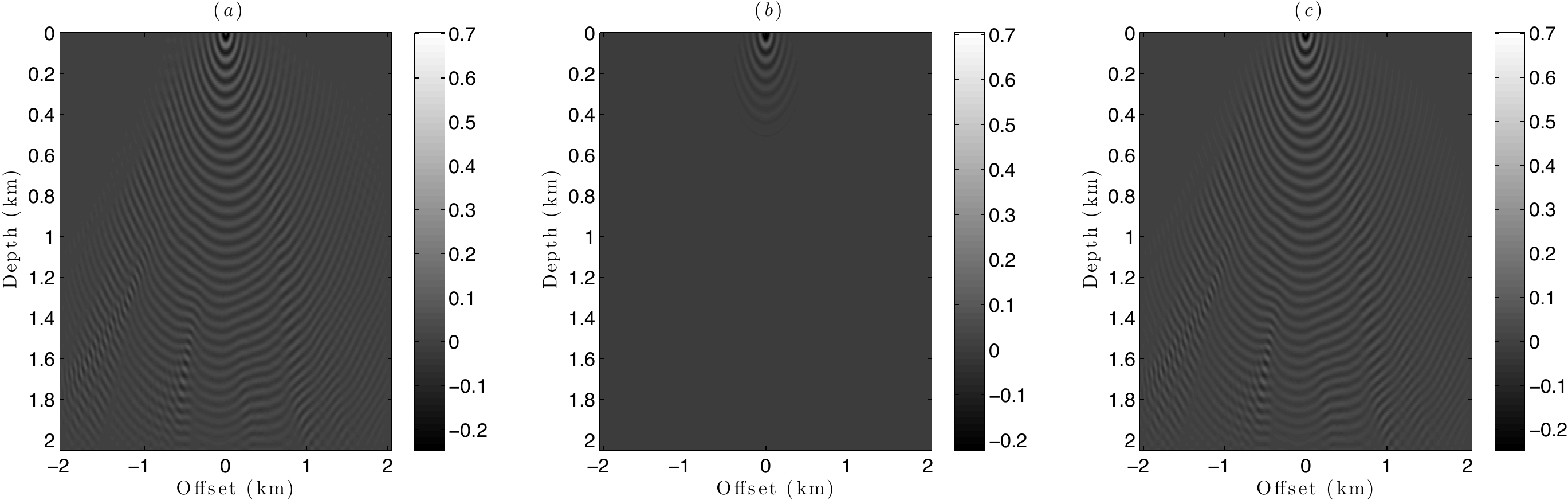}}
\caption{2D modelling results. a) modelling with LU factorization. b) GMRES without preconditioners. c) GMRES with shited-Laplace preconditioner ($\beta_1=1$ and $\beta_2=1$).}
\label{fig:6}
\end{figure}

\section{Conclusion}
We reported the use of shifted-Laplace preconditioners for the solution of heterogenous Helmholtz equation. For large scale problems, application of iterative algorithms such as GMRES method is inevitable. We showed that the vector support of Krylov subspace without the application of preconditioners is not enough to approximate the solution. The performance of GMRES algorithm was improved by introducing preconditioners to the system. The numerical results on 1D and 2D cases proved this claim. However, the degree of suitability of the preconditioners for the solution of inhomogeneous Helmholtz equation needs more research and analysis. 

\section{Acknowledgements}
The authors are grateful to the sponsors of Signal Analysis and Imaging Group (SAIG) at the University of Alberta.

\bibliography{paper.bib}

\end{document}